\documentclass[12pt,a4paper]{article}
\usepackage{a4wide}
\usepackage{amsmath}
\usepackage{amssymb}
\usepackage{epsfig}
\usepackage{subfigure}
\usepackage{exscale}
\usepackage{float}
\usepackage{bbm}
\usepackage[numbers,sort&compress]{natbib}
\usepackage{pst-plot, pstricks,pst-math}
\usepackage{fancybox,amssymb,color}
\usepackage{graphicx}
\usepackage{pstricks, color, graphicx, epsfig, psfrag}
\usepackage{amsfonts,amsmath,amssymb,slashed}
\usepackage{dsfont}
\usepackage{bbm,bm}
\usepackage{fancyhdr, a4wide}
\usepackage[english]{babel}
\usepackage{subfigure}

\makeatletter
\@addtoreset{equation}{section}
\makeatother

\title{Mermin-Wagner at the Crossover Temperature}

\author{Christoph P.\ Hofmann$^a$ \\ \\
\normalsize {$^a$ Facultad de Ciencias, Universidad de Colima} \\
\vspace{0.3cm}
\normalsize {Bernal D\'iaz del Castillo 340, Colima C.P.\ 28045, Mexico} \\}

\begin{document}

\maketitle

\begin{abstract} \normalsize

Mermin-Wagner excludes spontaneous (staggered) magnetization in isotropic ferromagnetic (antiferromagnetic) Heisenberg models at finite
temperature in spatial dimensions $d \le 2$. While the proof relies on the Bogoliubov inequality, here we illuminate the theorem from an
effective field theory point of view. We estimate the crossover temperature $T_c$ and show that, in weak external fields $H$, it tends to
zero: $T_c \propto \sqrt{H}$ ($d=1$) and $T_c \propto 1/|\ln H|$ ($d=2$). Including the case $d$=3, we derive upper bounds for the
(staggered) magnetization by combining microscopic and effective perspectives -- unfortunately, these bounds are not restrictive.

\end{abstract}

\maketitle

\section{Introduction}
\label{Intro}

In the article by Mermin and Wagner \citep{MW66}, absence of spontaneous symmetry breaking in isotropic Heisenberg models at finite
temperature in spatial dimensions $d \le 2$ is demonstrated by considering the magnetization or staggered magnetization in weak external
magnetic or staggered fields. When the external field tends to zero, while the finite temperature is kept fixed, the (staggered)
magnetization tends to zero as well. The theorem states that the functional dependence between (staggered) magnetization $m$ and weak
external field $H$ is characterized by a power law in $d$=1, while in $d$=2 the connection is logarithmic,
\begin{eqnarray}
\label{mw}
m & < & c_1 \, \frac{{H}^{1/3}}{T^{2/3}} \qquad (d=1) \, , \nonumber \\
m & < & c_2 \, \frac{1}{\sqrt{T} \sqrt{|\ln H|}} \qquad (d=2) \, .
\end{eqnarray}
Note that it is irrelevant whether ferromagnetic or antiferromagnetic interactions are considered.

The Mermin-Wagner theorem is based on a microscopic description of ferro- and antiferromagnets, and a crucial ingredient in its proof is
the Bogoliubov inequality \citep{Bog62} that turned out to be very useful in different contexts. In fact, the article by Hohenberg
\citep{Hoh67} on the absence of conventional superfluid or superconducting order in $d$=1 and $d$=2 is also based on the Bogoliubov
inequality.

Alternatively, systems exhibiting collective magnetic behavior can be analyzed within effective Lagrangian field theory. The question then
arises of how Mermin-Wagner reflects itself in the effective field theory point of view, and how effective and microscopic perspectives
are related to each other.

In analogy to the microscopic approach, we consider the (staggered) magnetization as a function of temperature and external field:
$m(T,H)$. For a given constant field strength $H$, the (staggered) magnetization decreases as temperature grows and eventually becomes
zero in the effective field theory description. We use the condition $m(T_c, H_c) = 0$ to estimate the crossover temperature $T_c$ in
terms of the external field. Although the effective theory operates at low temperatures, the extrapolation of the (staggered)
magnetization curves to the point $m=0$ still provides reasonable estimates for $T_c$.

In three spatial dimensions, $T_c$ tends to a finite value in the limit $H \to 0$: this defines the Curie (or N\'eel) temperature where
(anti)ferromagnetic order breaks down in a second order phase transition. Below $T_c$ -- in the absence of the external field --
spontaneous magnetic order exists. In lower spatial dimensions, however, the situation is qualitatively different: the crossover
temperature $T_c$, estimated from effective field theory, tends to zero in the limit $H \to 0$. Accordingly, no spontaneous magnetization
or spontaneous staggered magnetization can exist at finite temperatures in $d \le 2$. This is how the Mermin-Wagner theorem shows up in
the effective field theory description.

Remarkably, the functional dependence between $T_c$ and $H_c$ that we obtain from the condition $m(T_c, H_c) = 0$, is the same as in the
Mermin-Wagner inequalities (\ref{mw}): in $d$=1 (and $d$=3) we get a power law, $T_c \propto \sqrt{H_c}$, while in $d$=2 we find
$T_c \propto 1/|\ln H_c|$. Much like in the case of the Mermin-Wagner inequalities, the functional dependence is a consequence of the
spatial dimension only and does not depend on whether ferromagnetic or antiferromagnetic order is considered.

We further explore the interplay between effective and microscopic description by deriving upper bounds for the (staggered) magnetization
at the estimated crossover temperatures. Not only do we discuss the cases $d=1,2$, but also include ferromagnets and antiferromagnets in
three spatial dimensions. As it turns out, the upper bounds for the (staggered) magnetization are not restrictive. Nevertheless, we still
find it appropriate to briefly report our findings.

The rest of the paper is organized as follows. In Section \ref{merminWagner} we briefly recapitulate the essentials of the Mermin-Wagner
theorem, and then show how the theorem manifests itself in the effective field theory description. The connection between crossover
temperature and external field is derived in Section \ref{crossover} for ferromagnets and antiferromagnets in $d=1,2,3$. Upper bounds for
the (staggered) magnetization for the various systems of interest are established in Section \ref{bounds}. Finally, Section \ref{summary}
contains our conclusions.

\section{Mermin-Wagner Theorem}
\label{merminWagner}

\subsection{Rigorous Statement on the Microscopic Level}
\label{microscopicMW}

The theorem by Mermin and Wagner \citep{MW66} states that there can be no spontaneous symmetry breaking at finite temperature in the
isotropic Heisenberg model,
\begin{equation}
\label{HeisenbergModel}
{\cal H}_0 = - \frac{1}{2} \, \sum_{ij} J_{ij} \, {\vec S}_i \cdot {\vec S}_j \, ,
\end{equation}
in spatial dimensions less or equal two. The theorem includes both ferromagnetic ($J_{ij} > 0$) and antiferromagnetic ($J_{ij} < 0$) order.

More concretely, the authors consider the quantity $m(T,H)$: the (staggered) magnetization per particle as a function of temperature and
an external field $H$. In the case of the ferromagnet, this is the magnetic field that points into the $z$-direction,
\begin{equation}
{\cal H} = {\cal H}_0 - \sum_i S^z_i H \, ,
\end{equation}
while for antiferromagnetic coupling,
\begin{equation}
{\cal H} = {\cal H}_0 - \sum_i {(-1)}^i \, S^z_i H \, ,
\end{equation}
we are dealing with a staggered field.

In the limit $H \! \to \!0$, while keeping $T$ constant, the (staggered) magnetization tends to zero,
\begin{equation}
\lim_{H \to 0} m(T,H) = 0 \qquad (d \le 2) \, .
\end{equation}
Accordingly, spontaneous symmetry breaking is ruled out at finite temperature. The proof is based on the Bogoliubov inequality that leads
to the explicit relations
\begin{eqnarray}
\label{MWweakfield}
m & < & c_1 \, \frac{{H}^{1/3}}{T^{2/3}}\qquad (d=1) \, ,\nonumber \\
m & < & c_2 \, \frac{1}{\sqrt{T} {\sqrt{|\ln H|}}} \qquad (d=2) \, ,
\end{eqnarray}
provided that the external field $H$ is weak. Although the proof in the original article refers to the Heisenberg model, it can be
extended to the XY model or the Hubbard model, among others (see, e.g., Refs.~\citep{KLS81,Bru01,GN01,LPL11}). Furthermore, the analog of
the Mermin-Wagner theorem that emerges in relativistic field theories was first proven and discussed by Coleman in Ref.~\citep{Col73}.

\subsection{Manifestation of Mermin-Wagner on the Effective Level}
\label{eftMW}

The systems we address in this study are ferro- and antiferromagnetic films ($d$=2) as well as ferromagnetic spin chains ($d$=1). We also
include ferro- and antiferromagnetic crystals ($d$=3) in order to emphasize the qualitative difference with respect to the physics in lower
spatial dimensions. However, we do not consider antiferromagnetic spin chains, since they are more subtle both on the effective and
microscopic level.\footnote{Antiferromagnetic spin chains will be analyzed elsewhere.}

Complementary to the microscopic description where the Mermin-Wagner proof is based upon, we use effective field theory to explore the
low-temperature properties of ferro- and antiferromagnets. The method relies on the fact that the spin-waves -- the collective excitations
-- are the relevant degrees of freedom at low temperatures.\footnote{Pedagogical outlines of the effective Lagrangian method with
applications to condensed matter systems are, e.g., Refs.~\citep{Bur07,Bra10}.} The basic input we need is the dispersion relation of the
spin waves in the external field. Irrespective of the spatial dimension, the leading term for ferromagnetic spin waves is quadratic,
\begin{equation}
\label{disprelFerro}
\omega({\vec k},H) = \gamma {\vec k}^2 + H \, ,
\end{equation}
while the leading term for antiferromagnetic spin waves takes the relativistic form
\begin{equation}
\label{disprelAntiFerro}
\omega({\vec k},H) = \sqrt{v^2 {\vec k}^2 + \gamma_s H} \, .
\end{equation}
The external field ${\vec H}=(0,0,H)$ is aligned with the (staggered) magnetization vector ${\vec m}(T,H) = (0,0,m(T,H))$. The constants
$\gamma, \gamma_s$ and $v$ depend on the microscopic parameters $S$ (spin quantum number), $J$ (exchange integral), and $a$ (lattice
constant), as well as on the geometry of the system. Below, in the formulas for the (staggered) magnetization, we express the constants
$\gamma, \gamma_s, v$ in terms of microscopic parameters for each system under consideration.

With the dispersion relation we calculate the free energy density at one-loop order as
\begin{equation}
\label{zFree}
z = z_0 + n \, \frac{T}{{(2 \pi)}^d} \, \int d^dk \ln \Big( 1- e^{-\omega({\vec k},H)/T} \Big) \, ,
\end{equation}
where $z_0$ is the energy density of the vacuum and $n$ is the number of independent spin-wave excitations: in ferromagnets we have $n$=1,
in antiferromagnets we have $n$=2. The (staggered) magnetization is then obtained via
\begin{equation}
m(T,H) = - \frac{\partial z(T,H)}{\partial H} \, .
\end{equation}
For the various systems of interest we have \citep{Hof99b,Hof10,Hof12a,Hof13,Hof17}
\begin{itemize}
{\item Ferromagnetic spin chains
\begin{equation}
\label{ferro1D}
m(T,H) = S - \frac{1}{2 \pi^{1/2} J^{1/2} S^{1/2}} \, T^{1/2} \, \sum_{n=1}^{\infty} \frac{e^{- n H/T}}{n^{1/2}} \, .
\end{equation}}
\end{itemize}

\begin{itemize}
{\item Ferromagnetic films
\begin{equation}
\label{ferro2D}
m(T,H) = S - \frac{1}{4 \pi J S} \, T \, \sum_{n=1}^{\infty} \frac{e^{- n H/T}}{n} \, .
\end{equation}}
\end{itemize}

\begin{itemize}
{\item Simple cubic ferromagnetic crystals
\begin{equation}
\label{ferro3D}
m(T,H) = S - \frac{1}{8 \pi^{3/2} J^{3/2} S^{3/2}} \, T^{3/2} \, \sum_{n=1}^{\infty} \frac{e^{- n H/T}}{n^{3/2}} \, .
\end{equation}}
\end{itemize}

\begin{itemize}
{\item Antiferromagnetic films
\begin{equation}
\label{af2D}
m(T,H) = m_0 + \frac{m_0^{3/2} v}{8 \pi \rho^{3/2}} \, \sqrt{H}
+ \frac{m_0}{2 \pi \rho} \, T \, \ln \Big[ 1 - \exp\Big( - \frac{v m_0^{1/2}}{{\rho}^{1/2}} \frac{\sqrt{H}}{T} \Big) \Big] \, .
\end{equation}}
\end{itemize}

\begin{itemize}
{\item Simple cubic antiferromagnetic crystals\footnote{The quantity $\cal K$ involves so-called next-to-leading order effective constants
that are of order one, much like $\cal K$ itself.}
\begin{eqnarray}
\label{af3D}
m(T,H) & = & (S - \sigma) + \frac{{\cal K}}{64 \pi^2 S J} \, H
- \frac{1}{ 2 \sqrt{3} J^2 S^2} \, T^2 \, h_1(T,H) \, , \\
h_1(T,H) & = & \frac{6 J S}{\pi^2} \, \frac{H}{T^2} \, \int_0^\infty d \chi \,
\frac{\sinh^2{\chi}}{\exp \Big[ 2 \sqrt{3} {S}^{1/2} {J}^{1/2} \sqrt{H} \cosh{\chi}/T \Big] - 1} \, .\nonumber
\end{eqnarray}}
\end{itemize}

Although the above series have all been calculated up to three-loop order within effective field theory, only the leading one-loop
(non-interacting) contributions are required for the present study. The expressions for the ferromagnet are valid for arbitrary spin $S$
and the magnetization at $T$=0 corresponds to maximal spin alignment. As for the antiferromagnet, the exact ground state at $T$=0 does not
correspond to the configuration of maximal spin (anti)alignment, i.e., to the (hypothetical) N\'eel state. In $d$=2 we focus on
$S=\frac{1}{2}$ where very precise loop-cluster data is available -- not only for the square-lattice antiferromagnet, but also for the
square-lattice XY model that we include in our discussion. The numerical values for $m_0$ (staggered magnetization density at $T$=0),
$\rho$ (spin stiffness), and $v$ (spin-wave velocity) read \citep{WJ94,GHJNW09,GHJPSW11}:
\begin{eqnarray}
\label{loopClusterData}
& & m_0 = 0.30743(1) /a^2 \, , \quad \rho = 0.1808(4) \, J \, , \quad v = 1.6585(10) \, J a \qquad (\mbox{antiferromagnet}) \, ,
\nonumber \\ 
& & m_0 = 0.43561(1) /a^2 \, , \quad \rho = 0.26974(5) \, J \, , \quad v = 1.1347(2) \, J a \qquad (\mbox{XY model}) \, .
\end{eqnarray}
In $d$=3 we correct the N\'eel staggered magnetization at $T$=0 by the Anderson factor $\sigma$ which for the simple cubic lattice and for
$S=\frac{1}{2}$ is $\sigma=0.078$ \citep{And52}.

The above expressions that refer to three spatial dimensions are perfectly valid also in zero external field: the spontaneous (staggered)
magnetization is well-defined at finite temperatures. In lower spatial dimensions, on the other hand, the effective expansions for the
(staggered) magnetization become singular in the limit $H \to 0$, in accordance with the fact that spontaneous (staggered) magnetization
cannot exist in $d \le 2$ at finite temperatures. Let us analyze the physical implications of the limit $H \to 0$ on the effective level
in more detail.

To see the essential point, we first focus on $d$=3. For a given finite and constant value of the external field, the (staggered)
magnetization decreases as temperature grows. A characteristic quantity is the crossover temperature $T_c$ where the first derivative of
the (staggered) magnetization with respect to temperature develops a minimum. Then, if one approaches zero external field strength, the
situation becomes qualitatively different: in the limit $H \to 0$, the (staggered) magnetization eventually becomes zero at the critical
temperature -- the system undergoes a second order phase transition characterized by the Curie (N\'eel) temperature.

Now in the effective field theory description, we do not see these effects related to the crossover phase, because the effective framework
operates at temperatures low compared to the crossover or phase transition temperature. According to our effective formulas, the
(staggered) magnetization always falls to zero -- and even drops to negative values -- if temperature is raised sufficiently, signaling
that the effective field theory approach starts to break down. Still, imposing the condition $m(T_c,0)=0$ in our effective expansions,
provides us with an {\it estimate} for the Curie (N\'eel) temperature. For the simple cubic lattice and $S=\frac{1}{2}$ we get
\begin{equation}
T_C \approx 2.1 \, J \, , \qquad
T_N \approx 2.1 \, J \, .
\end{equation}
Comparing these values with the literature (see, e.g., the textbook by Kittel \citep{Kit96}), one concludes that the one-loop formulas
extrapolated to higher temperatures still yield accurate estimates for $T_C$ and $T_N$. This encourages using the same strategy in nonzero
external fields in order to obtain {\it estimates} for the crossover temperature $T_c$ from effective field theory through the condition
$m(T_c,H_c)=0$.

Let us turn to lower space dimensions. In $d \le 2$, for a given constant and nonzero value of the external field, again, the effective
curves for the (staggered) magnetization formally drop to zero at a temperature that defines our estimate for the crossover temperature
$T_c$. However, the crucial difference with respect to $d$=3 concerns the behavior in very weak external fields. In particular, in the
limit $H \to 0$, the quantity $T_c$ does not approach a finite value, but tends to zero. One concludes that a (nonzero) Curie or N\'eel
temperature does not exist in $d \le 2$ and that spontaneous (staggered) magnetization at finite temperatures does not emerge. This is how
the Mermin-Wagner theorem manifests itself on the effective level:
\begin{eqnarray}
\lim_{H \to 0} T_c & = & 0 \qquad (d \le 2) \, , \nonumber \\
\lim_{H \to 0} T_c & = & \{T_C, T_N\} \qquad (d=3) \, .
\end{eqnarray}
We emphasize that we are dealing with the observation of how Mermin-Wagner shows up in the effective field theory perspective. In no way
do we claim to have provided an alternative proof of absence of spontaneous symmetry breaking at finite $T$ in $d \le 2$.

\section{Strength of External Field at $T_c$}
\label{crossover}

The condition $m(T_c,H_c)=0$ relates the estimate for the crossover temperature to the value $H_c$ of the external field at $T_c$. With the
one-loop effective expansions we obtain
\begin{eqnarray}
\label{TcHc}
T_c & = & \alpha_1 \sqrt{H_c} \qquad (d=1) \, , \nonumber \\
T_c & = & \alpha_2 \frac{1}{|\ln H_c|} \qquad (d=2) \, , \nonumber \\
T_c & = & \{T_C, T_N\} + \alpha_3 \sqrt{H_c} \qquad (d=3) \, .
\end{eqnarray}
The coefficients $\alpha_d$ depend on microscopic parameters and the geometry of the system -- explicit expressions will be provided
below.

Interestingly, despite the fact that ferromagnets and antiferromagnets are quite different entities, the relations (\ref{TcHc}) are
universal: we find a power-law in one and three spatial dimensions, and logarithmic behavior in two spatial dimensions -- irrespective of
whether Heisenberg ferro- or antiferromagnets (or even XY-type models) are considered. Remarkably, the functional dependence between $T_c$
and $H_c$ is reminiscent of the functional dependence between $m$ and $H$ in the Mermin-Wagner inequalities (\ref{MWweakfield}): power law
in $d$=1 versus logarithm in $d$=2.

We now explore the universality of this functional connection in more detail. To that end we consider the (staggered)
magnetization $m=m(T,H)$,\footnote{The symbol ${\overline m_0}$ stands for the (staggered) magnetization at zero temperature,
${\overline m_0} = m(0,H)$, whereas the quantity $m_0$ represents the (staggered) magnetization at zero temperature in zero external
field, $m_0=m(0,0).$}
\begin{equation}
m = - \frac{\partial z}{\partial H} = {\overline m_0} - \frac{1}{{(2\pi)}^{d}} \, \int d^d k
\frac{\partial \omega/\partial H}{e^{\omega/T} -1} \, .
\end{equation}
The dispersion relations are given in Eq.~(\ref{disprelFerro}) and Eq.~(\ref{disprelAntiFerro}), respectively. Introducing spherical
coordinates, we get
\begin{eqnarray}
m & = & {\overline m_0} - \frac{T^{\frac{d}{2}}}{2^d {\pi}^{d/2} \Gamma(\frac{d}{2}) \gamma^{\frac{d}{2}}} \, \int_0^{\infty} dx
\frac{x^{\frac{d-2}{2}}}{e^{x+H/T} - 1} \qquad (\mbox{ferromagnet}) \, , \\
m & = & {\overline m_0} - \frac{\gamma_s T^{d-1}}{2^d {\pi}^{d/2} \Gamma(\frac{d}{2}) v^d} \, \int_0^{\infty} dx
\frac{x^{\frac{d-2}{2}}}{e^{\sqrt{x + \gamma_s H/ T^2}}- 1} \, \frac{1}{\sqrt{x + \gamma_s H/ T^2}} \qquad (\mbox{antiferromagnet}) \nonumber \, .
\end{eqnarray}
After some trivial manipulations, the integrals take the form
\begin{eqnarray}
m & = & {\overline m_0} - \frac{T^{\frac{d}{2}}}{2^d {\pi}^{d/2} \Gamma(\frac{d}{2}) \gamma^{\frac{d}{2}}} \,
\int_{\frac{H}{T}}^{\infty} dz
\frac{{(z-H/T)}^{\frac{d-2}{2}}}{e^z - 1} \qquad (\mbox{ferromagnet}) \, , \\
m & = & {\overline m_0} - \frac{\gamma_s T^{d-1}}{2^{d-1} {\pi}^{d/2} \Gamma(\frac{d}{2}) v^d} \, \int_{\frac{\sqrt{\gamma_s H}}{T}}^{\infty} dz
\frac{{(z^2-{\gamma_s H/ T^2})}^{\frac{d-2}{2}}}{e^z - 1} \qquad (\mbox{antiferromagnet}) \nonumber \, .
\end{eqnarray}
It is now clear that the difference between ferromagnetic and antiferromagnetic behavior is wiped out in two spatial dimensions. In weak
external fields, the condition $m(T_c,H_c)=0$ then implies the logarithmic connection\footnote{The effective constant $\gamma$ has been
expressed in terms of microscopic quantities as $\gamma = J S a^2$. This is valid in all dimensions $d=\{1,2,3\}$, see
Refs.~\citep{Hof12a,Hof13,Hof17}.}
\begin{eqnarray}
\label{tcD2}
t_c & = & \frac{4 \pi S^2}{|\log{h_c}|} \qquad (\mbox{ferromagnet}, \ d=2) \, , \nonumber \\
t_c & = & \frac{4 \pi \rho}{J} \, \frac{1}{|\log{h_c}|} \qquad (\mbox{antiferromagnet}, \ d=2) \, , \nonumber \\
t_c & = & \frac{8 \pi \rho}{J} \, \frac{1}{|\log{h_c}|} \qquad (\mbox{XY model}, \ d=2) \, .
\end{eqnarray}
Note that temperature and external field strength are measured in units of the exchange integral,
\begin{equation}
t = \frac{T}{J} \, , \qquad h = \frac{H}{J} \, .
\end{equation}

In one spatial dimension, the integral related to the ferromagnetic spin chain amounts to
\begin{equation}
m = {\overline m_0} - \frac{T^{\frac{1}{2}}}{2 \pi \gamma^{\frac{1}{2}}} \, \int_0^{\infty} dx \frac{x^{-\frac{1}{2}}}{e^{x+H/T} - 1}
= {\overline m_0} - \frac{T^{\frac{1}{2}}}{2 \pi^{\frac{1}{2}} \gamma^{\frac{1}{2}}} \, \sum_{n=1}^{\infty} \frac{e^{-n H/T}}{n^{\frac{1}{2}}} \, .
\end{equation}
In a weak external magnetic field one obtains the power law
\begin{equation}
\label{tcD1}
t_c = 2 S^{\frac{3}{2}} \sqrt{h_c} \qquad (\mbox{ferromagnet}, \ d=1) \, .
\end{equation}

In three spatial dimensions, the ferromagnet obeys 
\begin{equation}
m = {\overline m_0} - \frac{T^{\frac{3}{2}}}{4 \pi^2 \gamma^{\frac{3}{2}}} \, \int_0^{\infty} dx \frac{x^{\frac{1}{2}}}{e^{x+H/T} - 1}
= {\overline m_0} - \frac{T^{\frac{3}{2}}}{8 \pi^{\frac{3}{2}} \gamma^{\frac{3}{2}}} \, \sum_{n=1}^{\infty} \frac{e^{- n H/T}}{n^{\frac{3}{2}}}
\, .
\end{equation}
Accordingly, the connection between $t_c$ and $h_c$ in weak magnetic fields is
\begin{equation}
\label{dep3DF}
t_c = \frac{4 \pi S^{\frac{5}{3}}}{{ \{ \zeta(\frac{3}{2}) \} }^{\frac{2}{3}}}
+ \frac{8 \pi S^{\frac{5}{6}}}{{3 \{ \zeta(\frac{3}{2}) \} }^{\frac{4}{3}}} \, \sqrt{h_c} \qquad (\mbox{ferromagnet}, \ d=3) \, .
\end{equation}
In contrast to $d \le 2$, the expansion starts with a term that does not involve the external field. In the limit $h_c \to 0$, $t_c$
coincides with the estimate for the Curie temperature. For the antiferromagnet in three spatial dimensions, the relevant expression takes
the form
\begin{equation}
m = {\overline m_0} - \frac{\gamma_s T^2}{2 \pi^2 v^3} \, \int_0^{\infty} dk \frac{k^2}{e^{\sqrt{k^2 + \gamma_s H/ T^2}}-1}
\frac{1}{\sqrt{k^2 + \gamma_s H/ T^2}} \, .
\end{equation}
Following Ref.~\citep{ABS01}, the Taylor expansion of the function $J_1(\epsilon)$,
\begin{equation}
\label{J1}
J_1(\epsilon) = 8 \, \int_0^{\infty} dk \frac{k^2 }{\sqrt{k^2 + \epsilon^2}} \, \frac{1}{e^{\sqrt{k^2 + \epsilon^2}}-1} \, ,
\end{equation}
starts with
\begin{equation}
J_1(\epsilon) = \frac{4 \pi^2}{3} - 4 \pi {\epsilon} + {\cal O}(\epsilon^2 \log\epsilon) \, .
\end{equation}
In weak staggered fields, we thus have
\begin{equation}
\label{dep3DAF}
t_c = 2 \sqrt{2} 3^{\frac{3}{4}} S\sqrt{S-\sigma}
+ \frac{3 \sqrt{3} S^{\frac{1}{2}}}{\pi} \, \sqrt{h_c} \qquad (\mbox{antiferromagnet}, \ d=3) \, ,
\end{equation}
where the first contribution represents the estimate for the N\'eel temperature. Much like in $d$=2, the difference between ferromagnetic
and antiferromagnetic behavior is wiped out in three spatial dimensions: we have a square-root dependence between $t_c$ and $h_c$ both
in Eq.~(\ref{dep3DF}) and Eq.~(\ref{dep3DAF}).

\begin{figure}
\begin{center}
\includegraphics[width=6.0cm]{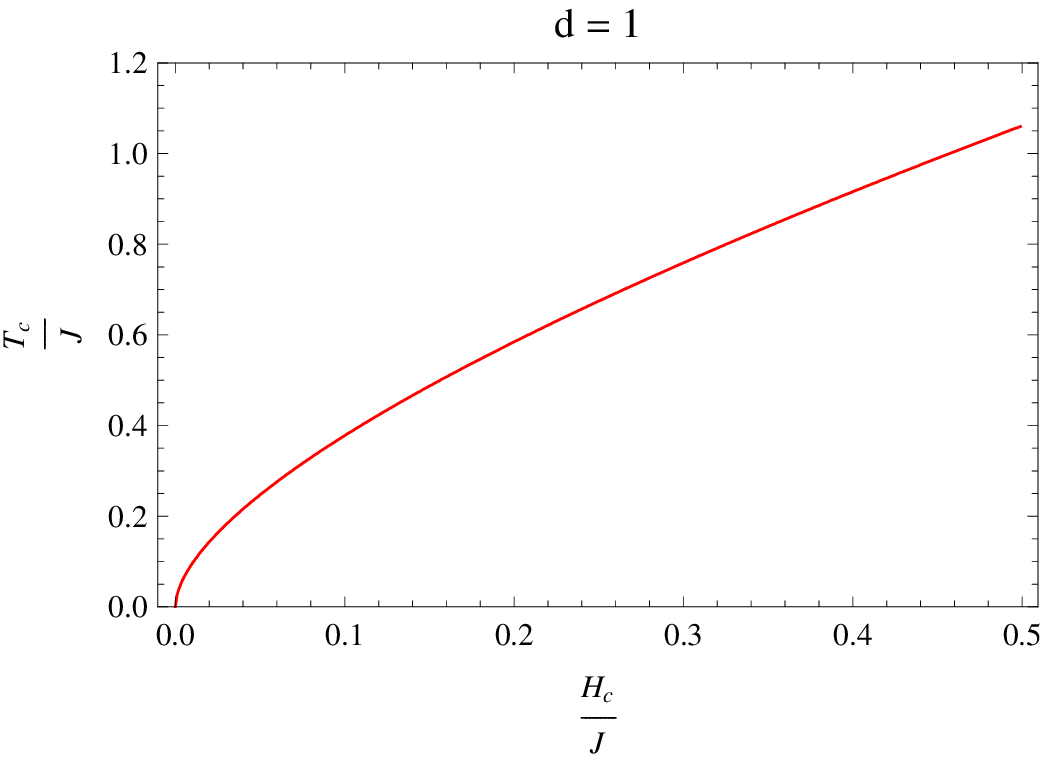} \includegraphics[width=6.0cm]{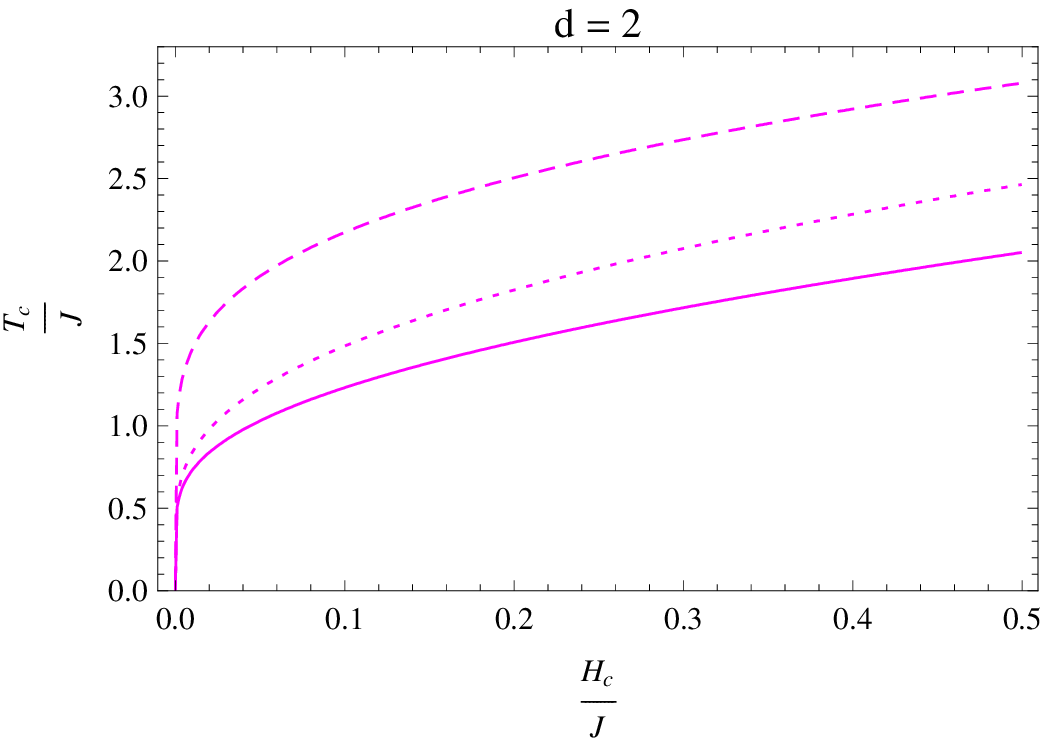} \\
\vspace{2mm}
\includegraphics[width=6.0cm]{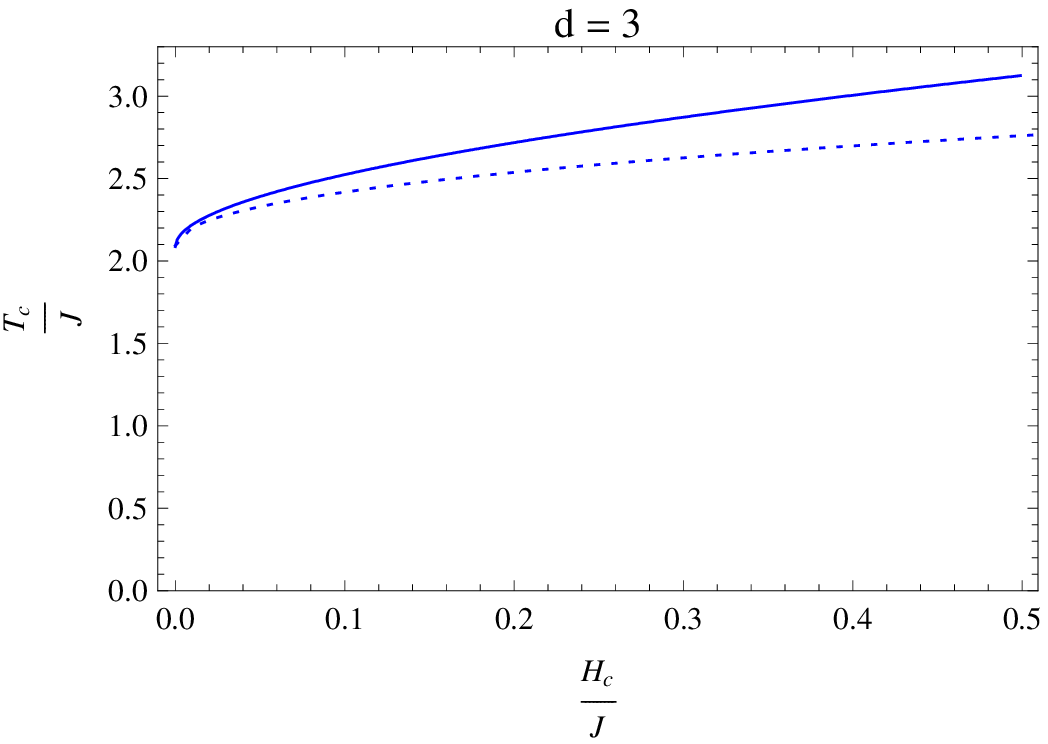} \\
\end{center}
\caption{[Color online] Estimated crossover temperatures $t_c=T_c/J$ versus external field $h_c=H_c/J$ for $S=\frac{1}{2}$ and for spatial
dimensions $d= 1,2,3$: ferromagnet (continuous), antiferromagnet (dotted) and XY model (dashed). In $d \le 2$, the quantity $t_c$ tends to
zero if the external field is switched off, while in $d$=3 it tends to the Curie (N\'eel) temperature.}
\label{figure1}
\end{figure}

Compiling previous results, the constants appearing in Eq.~(\ref{TcHc}) amount to
\begin{eqnarray}
\alpha^F_1 & = & 2 S^{\frac{3}{2}} J^{\frac{1}{2}} \, , \nonumber \\
\alpha^F_2 & = & 4 \pi S^2 J \, , \qquad \alpha^{AF}_2 = 4 \pi \rho \, , \qquad
\alpha^{XY}_2 = 8 \pi \rho \, , \nonumber \\
\alpha^F_3 & = & \frac{8 \pi S^{\frac{5}{6}} J^{\frac{1}{2}}}{{3 \{ \zeta(\frac{3}{2}) \} }^{\frac{4}{3}}} \, ,
\qquad \alpha^{AF}_3 = \frac{3 \sqrt{3} S^{\frac{1}{2}} J^{\frac{1}{2}}}{\pi} \, .
\end{eqnarray}
In Fig.~\ref{figure1} we illustrate the connection between $t_c$ and $h_c$ for the systems under consideration.

\section{Upper Bounds for the (Staggered) Magnetization}
\label{bounds}

In this section we combine effective and microscopic perspectives: what kind of information can be extracted from the Mermin-Wagner
theorem at the (estimated) crossover temperature? Before going into details, we have to take a closer look at how the Mermin-Wagner
inequalities are obtained. An intermediate step in the proof consists in replacing momentum sums by an integral,
\begin{equation}
\label{integrationMW}
S(S+1) > \frac{2 V_d \Omega_d m^2 T}{{(2 \pi)}^d} \int_0^{\bar k} \frac{dk k^{d-1}}{H m + S(S+1) {\bar J} k^2} \, ,
\end{equation}
where $\Omega_d$ is the surface of the $d$-dimensional unit sphere,
\begin{equation}
\Omega_d = d \, \frac{{\Big[ \Gamma(\frac{1}{2}) \Big]}^d}{\Gamma(\frac{d}{2}+1)} \, ,
\end{equation}
and $V_d$ is the volume per spin. To arrive at an analytical result, instead of integrating over the entire first Brillouin zone, in
Eq.~(\ref{integrationMW}) one only integrates over the (maximal) sphere with radius $\bar k$ that lies within the first Brillouin zone. In
one spatial dimension the integration leads to
\begin{equation}
\label{fullD1}
S(S+1) > \frac{2 T m^{3/2} a}{\pi \sqrt{{\bar J} S(S+1) H}} \arctan \Big( {\bar k} \sqrt{\frac{{\bar J} S(S+1)}{H m}} \Big)
\qquad (d = 1) \, ,
\end{equation}
while in two spatial dimension one gets
\begin{equation}
\label{fullD2}
S(S+1) > \frac{T m^2 a^2}{2 \pi {\bar J} S(S+1)} \ln \Big( 1 + \frac{{\bar J} S(S+1){\bar k}^2}{H m} \Big) \qquad (d = 2) \, .
\end{equation}
Restricting oneself to weak external fields, and inverting the above relations, one arrives at the Mermin-Wagner inequalities,
\begin{eqnarray}
\label{MWinequalities}
m & < & c_1 \, \frac{{H}^{1/3}}{T^{2/3}}\qquad (d=1) \, ,\nonumber \\
m & < & c_2 \, \frac{1}{\sqrt{T} \sqrt{|\ln H|}} \qquad (d=2) \, ,
\end{eqnarray}
with constants\footnote{Note that the value for $c_2$ refers to ferro- and antiferromagnets. In the case of the XY model we have
$c^{XY}_2 = \sqrt{2} c_2$. The explanation is given in the appendix.}
\begin{eqnarray}
c_1 & = & S(S+1) {\bar J}^{1/3} \, , \nonumber \\
c_2 & = & \sqrt{2 \pi} S(S+1){\bar J}^{1/2} \, .
\end{eqnarray}
Regarding $d$=2 we have assumed square lattice geometry (${\bar k} = \pi/a$). Convergence of the quantity $\bar J$,
\begin{equation}
\label{barJ}
{\bar J} = \frac{1}{2{\cal N}} \, \sum_{ij} \, |J_{ij}| \, {|{\vec x}_i - {\vec x}_j|}^2 \, ,
\end{equation}
ensures that the interaction described by the quantum Heisenberg model is of short range. $\cal N$ is the number of lattice sites.

We also include $d$=3, where the integration leads to
\begin{equation}
\label{fullD3}
S(S+1) > \frac{T m^2 a^3 {\bar k}}{2 \pi^2 {\bar J} S(S+1)} - \frac{T m^{5/2} a^3 \sqrt{H}}{2 \pi^2 {\bar J}^{3/2} S^{3/2}{(S+1)}^{3/2}}
\arctan \Big( {\bar k} \sqrt{\frac{{\bar J} S(S+1)}{H m}} \Big) \qquad (d = 3) \, .
\end{equation}
In the weak field limit one obtains for the simple cubic lattice (${\bar k} = \pi/a$):
\begin{equation}
\label{D3inequality}
m < c_3 \, \frac{1}{\sqrt{T}} \, , \qquad c_3 = \sqrt{\pi} S(S+1) {\bar J}^{1/2} \qquad (d = 3) \, .
\end{equation}

We insert the effective estimates for $t_c$ -- Eqs.~(\ref{tcD2}), (\ref{tcD1}), (\ref{dep3DF}), (\ref{dep3DAF}) -- into the microscopic
Mermin-Wagner inequalities, Eq.~(\ref{MWinequalities}), as well as into Eq.~(\ref{D3inequality}). For the various systems of interest, the
constants ${\cal C}_d$ in
\begin{equation}
\label{Tcinequalities}
m < {\cal C}_d
\end{equation}
are
\begin{eqnarray}
{\cal C}^F_1 & = & \frac{S+1}{2^{\frac{2}{3}}} {\Bigg( \frac{\bar J}{J} \Bigg)}^{\frac{1}{3}} \, , \nonumber \\
{\cal C}^F_2 & = & \frac{S+1}{\sqrt{2}} \, {\Bigg( \frac{\bar J}{J} \Bigg)}^{\frac{1}{2}} \, , \qquad
{\cal C}^{AF}_2 = \frac{S(S+1)}{\sqrt{2 \rho}} \, {\bar J}^{\frac{1}{2}} \, , \qquad
{\cal C}^{XY}_2 = \frac{S(S+1)}{\sqrt{2 \rho}} \, {\bar J}^{\frac{1}{2}} \, , \nonumber \\
{\cal C}^F_3 & = & \frac{S^{\frac{1}{6}}(S+1)}{2} \, { \{ \zeta(\mbox{$\frac{3}{2}$}) \} } ^{\frac{1}{3}} \,
{\Bigg( \frac{\bar J}{J} \Bigg)}^{\frac{1}{2}} \, ,
\qquad {\cal C}^{AF}_3 = \frac{\pi^{\frac{1}{2}} \sqrt{S}(S+1)}{2^{\frac{3}{4}} 3^{\frac{3}{8}} 4{(S - \sigma)}^{\frac{1}{4}}} \,
{\Bigg( \frac{\bar J}{J} \Bigg)}^{\frac{1}{2}} \, .
\end{eqnarray}
The inequalities (\ref{Tcinequalities}) provide upper bounds for the (staggered) magnetization at the crossover temperature. One should
keep in mind, however, that the analysis up to now refers to very weak external fields: the functional dependences between temperature
and external field in the Mermin-Wagner inequalities and in the estimates for $t_c$ are the same and therefore cancel in
(\ref{Tcinequalities}), such that the upper bounds reduce to the constants ${\cal C}_d$ that depend on microscopic parameters only. Let us
now discuss the general case where the external field is finite.

Unfortunately, with the framework established so far, the upper bounds for the (staggered) magnetization are not restrictive: it is a
priori obvious, e.g., that the magnetization of the ferromagnet can never be larger than the maximal value $S$. We now try to lower these
bounds, i.e., drive them into a domain where they become physically meaningful.

First, to arrive at simple analytical expressions, the integration in Eq.~(\ref{integrationMW}) is only performed over the (maximal)
sphere that fully lies within the first Brillouin zone. We now drop this idealization. In two spatial dimensions the integration over the
entire first Brillouin zone can still be performed analytically, but the corresponding expression is lengthy and not very illuminating.
Then, in $d$=3, the integration has to be performed numerically.

Second, we need to know the value of the quantity $\bar J$ defined by Eq.~(\ref{barJ}). A crude estimate can be obtained by summing over
nearest neighbors only, i.e., by setting any $J_{ij}$ not related to nearest neighbors to zero. In this case $\cal N$ corresponds to the
number of nearest neighbors of a given lattice site. Irrespective of the spatial dimension, we then obtain the estimate $\bar J = J a^2$,
where $a$ is the lattice constant. Note that we have assumed square and simple cubic lattice geometries. Of course, the question remains
how well actual (anti)ferromagnets are described by this crude implementation of short-range order in the Heisenberg model. For
illustrative purposes, in the following plots, we refer to this hypothetical scenario "nearest-neighbor interactions only".

We can now derive improved upper bounds for the (staggered) magnetization $m$ in finite external fields at the crossover temperature.
Again we combine the microscopic Mermin-Wagner inequalities with the effective equations for $t_c$ and solve for $m$. It should be pointed
out that the external field is now finite, both in the microscopic and the effective description -- the relevant expressions are
Eqs.~(2.10)-(2.14), as well as Eqs.~(\ref{fullD1}), (\ref{fullD2}), and (\ref{fullD3}).

\begin{figure}
\begin{center}
\includegraphics[width=8.5cm]{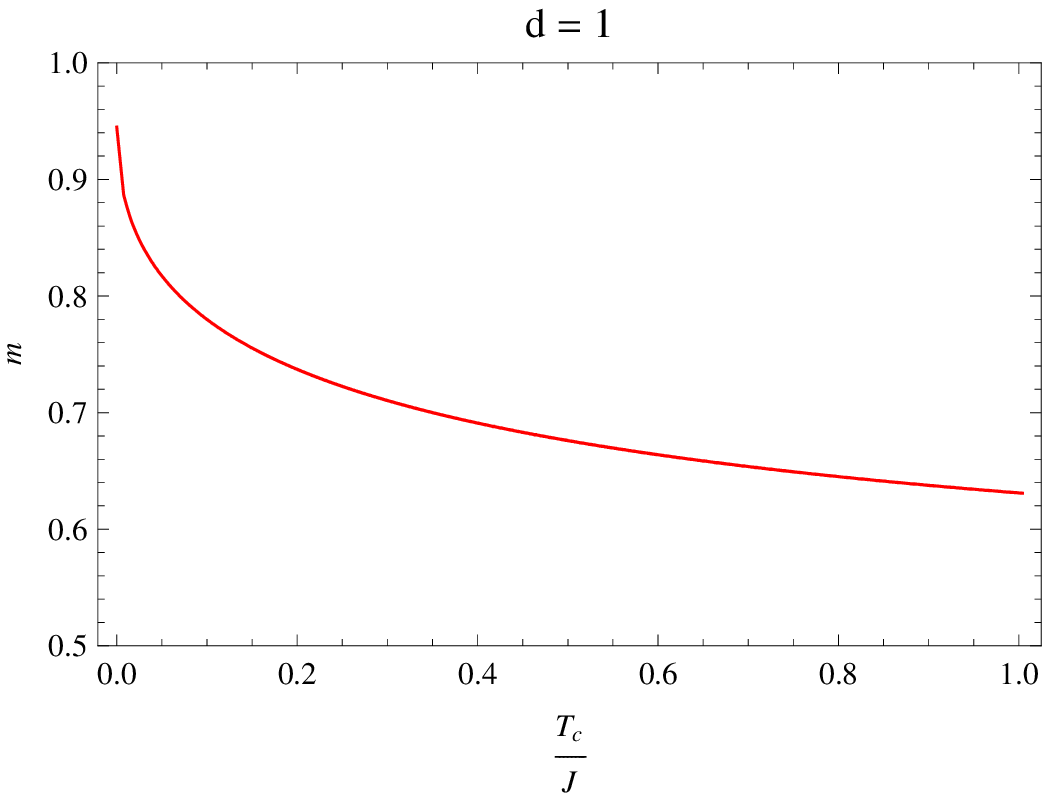} \\
\vspace{2mm}
\includegraphics[width=8.5cm]{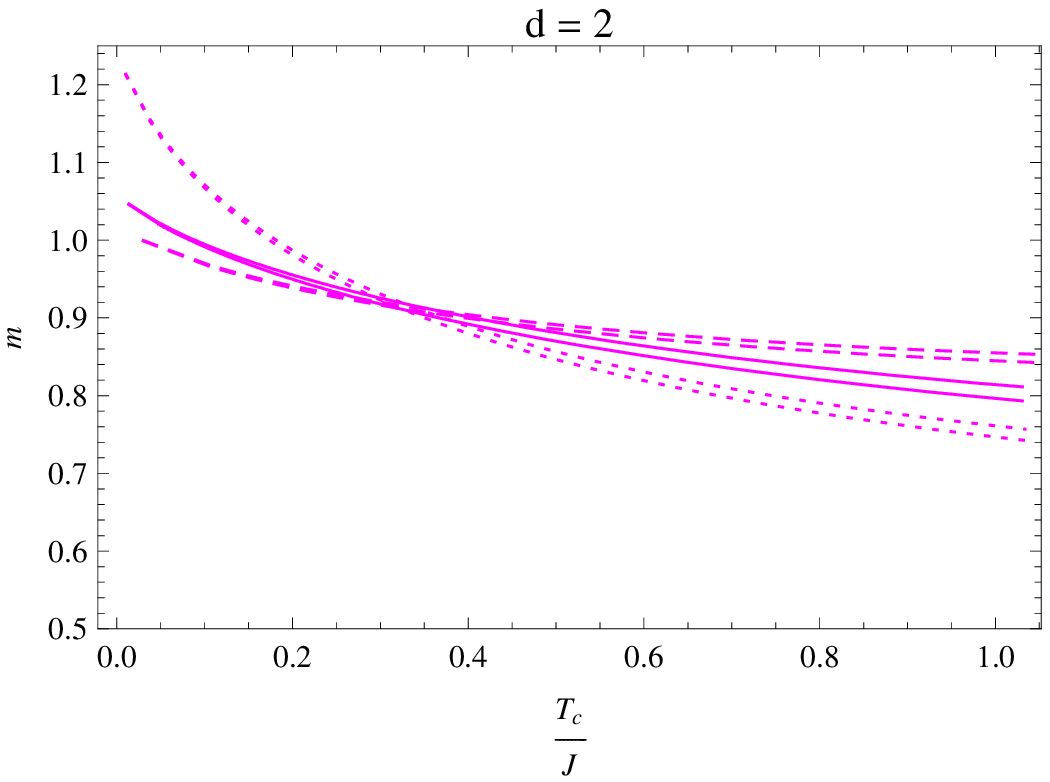} \\
\vspace{2mm}
\includegraphics[width=8.5cm]{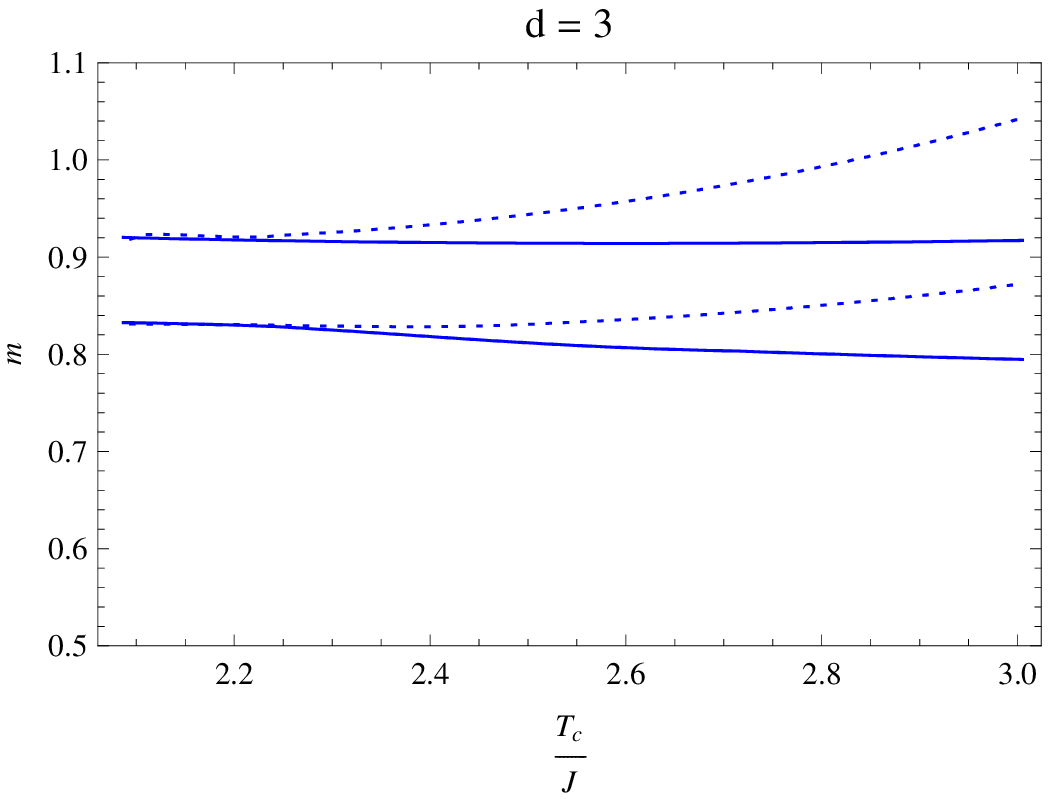} \\
\end{center}
\caption{[Color online] Upper bounds for the (staggered) magnetization $m$ at the crossover temperature $t_c=T_c/J$: ferromagnets
(continuous), antiferromagnets (dotted), and XY model (dashed), in spatial dimensions $d= 1,2,3$, and for $S=\frac{1}{2}$.}
\label{figure2}
\end{figure}

The resulting curves that all refer to $S=\frac{1}{2}$, are depicted in Fig.~\ref{figure2}. Continuous curves refer to the ferromagnet,
dotted curves to the antiferromagnet, and dashed curves to the XY model. In $d$=2 and $d$=3 we depict two situations for each system:
integrating over the entire first Brillouin zone versus integrating over the sphere with radius $\bar k$ lying inside the first Brillouin
zone. As witnessed by the figures, integrating over the entire Brillouin zone leads to lower upper bounds -- while the difference is small
in $d$=2, it is more pronounced in $d$=3.

Unfortunately, the improved bounds in zero and nonzero external fields still turn out to be larger than the maximal ($T$=0) values for the
(staggered) magnetization. The outcome is therefore negative: on the basis of the Mermin-Wagner inequalities, it is not possible to
extract reasonable (physically restrictive) upper bounds for the (staggered) magnetization. After all, the Bogoliubov inequality is
designed to demonstrate absence of spontaneous symmetry breaking in zero external field, but not to derive upper bounds for the
(staggered) magnetization.

\section{Conclusions}
\label{summary}

The effective field theory description of ferromagnets and antiferromagnets relies on the observation that the spin waves are the
relevant low-energy degrees of freedom. Extrapolating the one-loop effective expansions for the (staggered) magnetization $m(T,H)$
to higher temperatures, the condition $m(T_c,H_c)=0$ provides an implicit relation between the crossover temperature $T_c$ and the
external field at $T_c$. We have used this strategy to estimate the crossover temperature for ferromagnets and antiferromagnets in one,
two, and three spatial dimensions. 

First, there is a qualitative difference between $d$=3 and $d \le 2$, respectively. In three spatial dimensions, $T_c$ tends to a finite
value if the external field is switched off: this is the Curie (N\'eel) temperature where spontaneous magnetic order disappears in a
second order phase transition. On the other hand, in lower space dimensions, $T_c$ tends to zero in the limit $H \to 0$ -- consistent with
the fact that spontaneous magnetic order does not exist in spatial dimensions $d \le 2$ at nonzero temperatures. This is how the
Mermin-Wagner theorem emerges in the effective field theory description.

Second, the functional dependence between $T_c$ and $H_c$ resulting from the condition $m(T_c,H_c)=0$, is universal in weak external
fields: in $d$=1 and $d$=3, we have a power law, whereas in $d$=2 the connection is logarithmic -- irrespective of whether ferromagnetic
or antiferromagnetic coupling is considered. This is reminiscent of the universality of the functional dependence between temperature and
external field in the Mermin-Wagner inequalities: the power law in $d$=1, and the logarithmic behavior in $d$=2, are a consequence of the
spatial dimension.

We then have combined the effective description with the microscopic perspective that underlies the Mermin-Wagner theorem, trying to
extract upper bounds for the (staggered) magnetization at the estimated crossover temperature. It should be noted that the logic of the
present study is different from the one inherent in the article by Mermin-Wagner. We do not aim at proving absence of spontaneous symmetry
breaking (i.e., analyzing the limit $H \to 0$ while keeping $T$ finite), but use the Mermin-Wagner inequalities to derive upper bounds for
the (staggered) magnetization in presence of a {\it finite} external field. Unfortunately, the upper bounds we obtain are not restrictive.

\section*{Acknowledgments}
The author thanks A.\ Auerbach, N.\ D.\ Mermin, and U.-J.\ Wiese for correspondence and stimulating discussions.

\begin{appendix}

\section{Heisenberg Model versus XY Model}
\label{appendix}

An intermediate step in the Mermin-Wagner proof consists in analyzing the inequality\footnote{See, e.g., Eq.~(6.29) in the textbook by
Auerbach \citep{Aue98}.}
\begin{equation}
\label{expvalue}
{\cal N}^{-1} \sum_i \langle {(S^y_i)}^2 \rangle \le S(S+1) \, ,
\end{equation}
where $\cal N$ is the total number of lattice sites. Since the external field points into the $z$-direction, the system must be isotropic
with respect to the $x$- and $y$-directions, i.e., the following expectation values are the same,
\begin{equation}
\sum_i \langle {(S^x_i)}^2 \rangle = \sum_i \langle {(S^y_i)}^2 \rangle \, .
\end{equation}
In general, the inequality
\begin{equation}
{\cal N}^{-1} \Big( \sum_i \langle {(S^x_i)}^2 \rangle + \sum_i \langle {(S^y_i)}^2 \rangle \Big) \le S(S+1)
\end{equation}
holds, and therefore in Eq.~(\ref{expvalue}) we gain a factor of one half, 
\begin{equation}
{\cal N}^{-1} \sum_i \langle {(S^y_i)}^2 \rangle \le \frac{1}{2} \, S(S+1) \, ,
\end{equation}
provided that we are referring to the Heisenberg model. Concerning the XY model, we have to rely on the estimate (\ref{expvalue}). 

\end{appendix}

\end{document}